\documentclass[aps,preprint,showpacs,amsmath,pre,amssymb]{revtex4-1} 
\usepackage{graphicx,dcolumn,bm,bbm,amssymb}
\usepackage{times}

\begin{document}

\hfill{\it Nature Materials 11, 608-613 (2012)}

\medskip

\centerline{\bf\large Mechanical Metamaterials with Negative Compressibility Transitions}

\vspace{0.4cm}

\centerline{Zachary G. Nicolaou$^{1}$ and Adilson E.\ Motter$^{1,2*}$} 

\baselineskip 14pt

{\small $^1$Department of Physics and Astronomy, Northwestern University, Evanston, Illinois 60208, USA, $^2$Northwestern Institute on Complex Systems, Northwestern University, Evanston, Illinois 60208, USA. $^*$e-mail: motter@northwestern.edu.}

\vspace{0.6cm}

\noindent
{\bf 
When tensioned, ordinary materials expand along the direction of the applied force.  Here, we explore network concepts to design metamaterials exhibiting negative compressibility transitions, during which a material undergoes contraction when tensioned (or expansion when pressured). 
Continuous contraction of a 
material in the same direction of an applied tension, and in response to this tension, is inherently unstable. 
The conceptually similar effect we demonstrate can be achieved, however, through destabilisations of (meta)stable equilibria of the constituents. These destabilisations give rise to a stress-induced solid-solid phase transition associated with a twisted hysteresis curve for the stress-strain relationship. The strain-driven counterpart of negative compressibility transitions is a force amplification phenomenon, where an increase in deformation induces a discontinuous increase in response force. We suggest that the proposed materials could be useful for the design of actuators, force amplifiers, micro-mechanical controls, and protective devices.
}

\vspace{0.4cm}

Metamaterials, engineered materials that gain their properties from structure rather than composition, have led to the study of a myriad of properties not exhibited by most (and in some cases all) natural materials. These new properties are typically characterised by negative constants, moduli, or indexes. In the case of electromagnetic properties, a revolution started a decade ago with the experimental realisation of Veselago's predictions \cite{Veselago68} on hypothetical materials with negative refractive index \cite{Smith00,Shelby01}, and related applications such as to the design of sub-wavelength lens \cite{Pendry00} and electromagnetic cloaks \cite{Schurig06}. Similar concepts have also been explored to create materials with unusual acoustic properties, including negative effective mass density for acoustic focusing \cite{Zhang09}. These properties contrast with those of most conventional materials. 

Conventional materials are also known to deform along the direction of an applied force in such a way that they expand when the force is tensional and contract when it is compressive, regardless of how they respond in transverse directions. Thus, one can ask whether a material could be designed to exhibit the opposite behaviour, corresponding to negative compressibility.
Previous studies have shown that porous materials can expand under hydrostatic pressure by sorption or infiltration of particles of the medium \cite{Baughman98,Lee02,Vakarin06,grima08,gatt08} and that resonant composites can exhibit negative effective elastic constants in response to acoustic waves \cite{Liu00,Fang06,Sheng03}. Similarly,  it has been shown that one component of a composite material (but not the material itself) can exhibit decreasing response force as a function of increasing applied deformation  \cite{Lakes01}. These negative responses have led to important applications, such as composites with unusually large stiffness \cite{jaglinsk07} or acoustic shielding potential \cite{Liu00}, but they rely on the system being either mechanically open or dynamically excited. The question being raised here, however, is whether the material itself (a thermodynamically closed system) can exhibit decreasing response deformation as a function of increasing applied force (changed over macroscopic timescales).

The fundamental problem with this question is that 
closed systems are thermodynamically forbidden from expressing such negative compressibility \cite{Reichl2009}.  
Indeed, for a closed system in equilibrium, if increase in force leads to deformation opposing the applied force, then the original configuration is necessarily unstable. This has been interpreted as an indication that negative compressibility in the direction of the applied forces cannot be realised for any closed natural or engineered material. However, this argument tacitly assumes that the changes in force are small and that the  
equilibrium survives this change. While true for infinitesimal changes, here we show that this is not necessarily true if the force is changed by a finite amount, as expected in many realistic situations. In this case, the equilibrium can become unstable or disappear, forcing the system to move to a different equilibrium. We exploit this mechanism to design and theoretically analyze materials that have this property, and for which the new equilibrium corresponds to a contraction (dilation) when the system is tensioned  (pressured), as illustrated in Fig.\ \ref{fig1}a,b.  
We refer to this discontinuous change as a {\it negative compressibility transition}. This transition manifests itself as an infinite negative value of the differential compressibility, which is also suitably expressed as a finite negative value of the finite-difference compressibility (see Table S1 for a summary of definitions).
Note that this is a bulk property, since volume can decrease (increase) under increased tension (pressure) due to the material's negative response along each and every direction of the applied external force. 

Negative compressibility transitions are distinct from negative Poisson's ratio  \cite{Lakes87,Baughman98-2} and negative normal stress \cite{Janmey07}, where the unusual response is transverse to the applied force.    They are also distinct from  negative incremental stiffness \cite{Lakes01,Moore06}, characterised by a decrease in the resulting restoration force for increasing deformation. Finally, the proposed negative compressibility  should be contrasted with  stretch densification and the related phenomena of linear and area expansion induced by hydrostatic pressure in closed systems \cite{Baughman98,Goodwin08,Fortes2011}, which are compensated by contraction in other directions, leading to ordinary behaviour for the bulk modulus. 
Figure \ref{fig1}c-e contrasts negative compressibility transitions with negative Poisson's ratio, negative incremental stiffness, and stretch densification.

In order to demonstrate that materials can exhibit negative compressibility transitions, we consider constituent elements formed by a system of four particles separated by distances $x$, $y$, $z$ and $h$ and interacting via general potentials $V_x$, $V_y$, $V_z$ and $V_h,$ as indicated in Fig.\ \ref{fig2}a. This system is characterised by the total potential
\vspace{-0.10cm}
\begin{equation}
V(x,y,h,F)=V_x(x)+V_y(y)+V_z(y-x)+V_x(h-y)+V_y(h-x)+V_h(h)-Fh,
\label{mec1}
\vspace{-0.10cm}
\end{equation}
where $F$ is an externally applied force (positive for tension and negative for pressure).
We also assume there is energy dissipation determined by contact with a heat bath, so that the system can reach equilibrium after being perturbed. As in the case of all previously studied closed systems, negative compressibility cannot occur continuously (Methods Summary). However, negative compressibility transitions are still possible in this system by exploiting a bifurcation similar to the one that occurs for the potential $U(\xi, {\cal F})=-\xi^3/3+{\cal F}\xi$, where $\xi$ is a variable and ${\cal F}$ an external parameter. For ${\cal F}>0$, this potential has a stable equilibrium point at $\xi^*=-\sqrt{\cal F}$ and an unstable one at $\xi^*=\sqrt{\cal F}$; for ${\cal F}=0$ there is a single degenerate 
equilibrium point at $\xi^*=0$; for ${\cal F}<0$ there is no equilibrium point. Therefore, if ${\cal F}$ is decreased from positive through zero, the stable equilibrium point vanishes giving rise to a destabilisation, and the system responds discontinuously. 

In the case of our system, it is relevant to destabilise the $V_z$ bond by increasing (decreasing) the force $F$ so as to drive the system to a different equilibrium corresponding 
to smaller (larger) $h$. This can be shown to be possible for potentials satisfying suitable conditions. 
We assume $F$ to be a tensional force, as reciprocal results can be achieved for pressures. 

The destabilisation weakens the $V_z$ bond and causes the system to transition from a predominantly series $V_x$---$V_z$---$V_x$  configuration (which we call the {\it coupled} configuration) to a predominantly parallel $V_x$---$V_y$, $V_y$---$V_x$ configuration (the {\it decoupled} configuration). This causes the bonds defined by $V_x$ to contract as the stress becomes distributed between them. If the shortening of $x$ overcomes the lengthening of $y$, the system as a whole contracts (i.e., $h$ decreases) in response to increasing tension (Fig.\ \ref{fig2}a).  This is achieved if the  potential $V_x$ is ``soft'' in the neighbourhood of  the transition 
but ``hardened'' for smaller $x$ and/or the  potential $V_y$ is ``soft''  in the neighbourhood of transition 
but ``hardened" for larger $y$.  This both illustrates that harmonic potentials cannot lead to the desired behaviour in  such one-dimensional systems (although they can in two-dimensional configurations---see Supplementary Information, Fig.\ S2) and  provides a criterion for the selection of the potentials (see Methods Summary).

Our design of this system is inspired by the key observation that the realised 
equilibrium is not necessarily optimal in a decentralised network. The best known precedent to this is the insight from Braess \cite{Braess05} that adding a road to a traffic network may increase rather than decrease the average travel time. Related concepts have been considered in connection with computer science and game theoretical studies \cite{Roughgarden05}, and have long been part of the transportation \cite{Beckmann56} and economics \cite{Pigou20} literature, amounting to over 90 years of multidisciplinary research. Analogous effects have also been identified in physical networks, including increase of current upon the removal of an intermediate conductor in electric networks and increase of restoring force upon the removal of a support string in spring-string networks \cite{Cohen91}. All these are examples in which the equilibrium realised by the system can be brought closer to the optimum by constraining the structure of the network. Our system is devised such that a conceptually similar phenomenon occurs {\it spontaneously}, in response to a change in 
the external parameter $F$
rather than in the structure of the network. 

Negative-compressibility metamaterials can be designed using this system as constituent elements, as shown in Fig.\ \ref{fig2}b for a square lattice configuration.  We first consider this material in the limit of small temperatures, $T$, and for changes of the external force sufficiently slow that the lattice is in quasi-static equilibrium at any particular time (except during destabilisation transitions). Under these conditions, the material exhibits a hysteresis loop for stress-driven cycles that, as shown in Fig.\ \ref{fig2}c, is identical to the one predicted for the constituents. 
Starting from a coupled configuration for all constituents (bottom left), increase in tensional stress (denoted $\sigma$ and defined as tension per lattice spacing) leads to increase in deformation, $\varepsilon$, until a decoupling destabilisation occurs at $\sigma=\sigma^d$. A negative compressibility transition is observed at this point, where the material contracts by a finite amount (i.e., $\varepsilon$ decreases) upon an arbitrarily small increase of $\sigma$.   This discontinuous behaviour is indicative of a first-order phase transition in the material.  The finite change in strain can be regarded as a ``latent strain,'' broadly analogous to the latent heat absorbed during, say, the vaporization of water.  Stress in our system plays the role of the temperature in this liquid-vapour analogy.  The negative compressibility in our system is then the result of a {\it negative latent strain}, i.e., a finite drop in $\varepsilon$ upon a small increase in $\sigma$.  After the decoupling, further increase in $\sigma$ leads to new increase in $\varepsilon$. More interesting, however, is the effect of reducing $\sigma$ after the transition.  This eventually leads to a coupling destabilisation at $\sigma^c$ (analogous to the condensation of water), which is also discontinuous, closing the hysteresis  loop.  For the coupling transition, the latent strain is positive, as a finite drop in $\varepsilon$ results from a small decrease in $\sigma$, making this a positive compressibility transition.  Each of these directional transitions is analogous to nonequilibrium phase transitions occurring, for example, in the Schl\"ogl chemical reaction model \cite{Reichl2009}, and they can be interpreted as different halves of the same phase transformation, which are set apart by hysteresis. For an animation of the response of the material to different stress profiles, see  Supplementary Movie.     

These destabilisations can be related to spinodals (points where metastable phases cease to exist), which have been considered, for example, in the stability limit hypothesis for the anomalous behaviour of water \cite{stanley}. The possibility of destabilisations occurring on a microscopic scale and resulting in a phase transition has also been examined in the context of ferroelastics,  such as for martensitic transformations and shape memory effects \cite{otsuka}, and theoretically in continuum mechanics \cite{abeyaratne} and discrete systems \cite{Puglisi2000}.  However, the possibility of materials exhibiting negative compressibility transitions like those considered here has not been previously recognised.

A complementary but important aspect concerns the response of the material when the deformation $\varepsilon$ is taken as the tunable external parameter. This defines a hysteresis loop for strain-driven cycles, which is shown in Fig.\ \ref{fig2}d. 
Ordinarily, when the (thermodynamic conjugate variable) strain is taken as the controlled variable, 
the stress remains constant during the phase transformation as the latent strain is overcome through the formation of a phase mixture.
In our system,
such behaviour does occur during the coupling transition, but during the decoupling transition, the latent strain is negative.
In this case, once the applied strain is increased past the destabilisation at  $\varepsilon^d$, the resulting stress $\sigma$ must increase discontinuously to become compatible with the new strain, giving rise to a mechanism for force amplification.  Moreover, instead of developing a phase mixture in which the proportions change continuously, the only possibility for this process is that the entire system (or a finite fraction of it, depending on how the strain is applied)  undergoes the decoupling transition concurrently. We note that the conclusions from Fig.\ \ref{fig2}c,d are valid for a cycling time $\tau$ (the time to vary the stress and strain between their minimum and maximum values)  larger than the characteristic time scale $\tau_0$ for the system to approach equilibrium in the limit of small $T$.

Figure \ref{fig3} shows how the phenomenology changes for finite temperatures. For vanishingly small $T$, the thermodynamic equilibria reduce to the configurations that are local minima of the potential, as considered in Fig.\ \ref{fig2}.  As the temperature is increased from zero, the initially stable configuration around a local minimum becomes metastable. 
But how does hysteresis, and hence the rate at which the thermal fluctuations allow the system to escape a local minimum of the free energy, depend on the temperature?  To address this question,  we established a rigorous model to estimate the effective energy barrier for both decoupling ($E^c_b$) and coupling ($E^d_b$) transitions (Supplementary Information).
While the decay of metastability depends on the free energy,  it is the height of an effective energy barrier---which also accounts for the frequency with which the free energy barrier is approached, the cycling time, and the size of the sample---that determines the transitions to the global minimum (Supplementary Information). For stress cycles, for example, the effective energy barrier for decoupling transitions in a square lattice of $N\times N$ constituents 
satisfies
\begin{eqnarray}
\label{couplebarrier}
E_b^c(\sigma,N,\tau)>\frac{2V_b^c(\sigma)}{\log{\left[\displaystyle\frac{N^2}{\pi}\frac{\sigma-\sigma^c}{\sigma^d}\tau\left(\omega_L(\sigma)+\omega_T(\sigma)\right)\right]}} \, ,
\label{eq_Eb}
\end{eqnarray}
where  $V_b^c$ is a potential barrier, and $\omega_L$ and $\omega_T$ are the frequencies of the lattice vibration modes that
give rise to decoupling events.
To test this model, we have implemented molecular dynamics simulations (Methods Summary)  for both stress cycles (Fig.\ \ref{fig3}a,c) and strain cycles (Fig.\ \ref{fig3}b,d). In all cases, the transitions occur earlier than in the zero temperature limit. More important, the onset of the transitions (Fig.\ \ref{fig3}c,d) is determined by the points at which the thermal energy $k_BT$ becomes comparable to the effective energy of the barrier,  $E^{c,d}_b$ (Fig.\ \ref{fig3}a,b). The transitions are determined using a cycling time $\tau$  larger than $\tau_0$. For nonzero $T\ll E^{c,d}_b/k_B $, $\tau_0$  can be interpreted as a time scale for the metastable state to be approached.

The evolution of the phase transition at nonzero temperatures depends on the interplay between three time scales: the cycling time $\tau$, the characteristic time to approach metastable equilibrium $\tau_0$, and the characteristic time to approach thermodynamic equilibrium, $\tau_t$. In general, $\tau_0$ is larger than but comparable to the time scale for volume fluctuations $\tau_V$, and $\tau_t$ is larger than but comparable to the time scale for transitions across the energy barrier.  Consider, for example, the stress cycles shown in Fig.\ \ref{fig3}a,c.  The coupled equilibrium of the potential  has lower energy for small stresses, while the decoupled equilibrium has lower energy for sufficiently high stresses. At some critical stress, the energy of the two equilibria coincide.  Equilibrium statistical mechanics 
predicts that the phase transition would proceed at this critical stress. 
Such an equilibrium phase transition cannot be a negative compressibility transition, as examination of the differential change in free energy at fixed temperature reveals that the high-pressure phase is always also the low-volume phase. However, the occurrence of an equilibrium phase transition involves the assumption that stress is changed arbitrarily slowly. For a fixed but nonzero rate of change (characterised by $\tau$), the external parameter may change before the system has approached its thermodynamic equilibrium (characterised by $\tau_t$) or even metastable equilibrium (characterised by $\tau_0$). If $\tau$ is larger than but comparable to $\tau_0$, it follows from the Boltzmann distribution that the temperature-dependent time scale $\tau_t$ will be exponentially large for $k_BT \ll E^{c,d}_b$, 
while $\tau_t$ becomes comparable to or smaller than $\tau$ as $k_BT$  becomes comparable to $E^{c,d}_b$.   
This underlies much of the properties shown in Fig.\ \ref{fig3} and is expected to be representative for laboratory time scales.  Larger temperatures and longer cycle times cause the metastable states to decay earlier in the cycle, but our molecular dynamics simulations show that 
transitions giving rise to
negative compressibility and force amplification persist for a wide range of small temperatures  and for large cycle times (Supplementary Information, Figs.\ S3 and S4).  Furthermore,  as indicated in Eq.\ (\ref{eq_Eb}), we can show that $E^{c,d}_b$ depends only logarithmically on the cycling time $\tau$ and size $N$ of the sample, indicating that orders of magnitude changes in them produce only moderate changes in the effective energy barrier (Supplementary Information).

This last observation is a very important. We conservatively predict that systems with typical atomic-sized length and energy scales can exhibit negative compressibility transitions in macroscopically accessible time scales.  For example, given an energy barrier typical of a covalent bond, $\Delta {V_b}^c \sim 1.5$ eV, microscopic frequencies typical of atomic vibrations, $\omega_L,\omega_T \sim 10^{14}$ s$^{-1}$, and macroscopic system size, $N^2 \sim 10^{23}$, we find that the effective energy barrier $E_b^c$ becomes comparable to thermal fluctuations at room temperature, $k_BT = 0.026$ eV, only when $\tau \gtrsim 10^{13}$ s. Thus, metastable states with such an energy barrier will survive thermal fluctuations for over $3 \times 10^5$ years!  But of course $\Delta {V_b}^c$ decreases in size as $\sigma$ approaches $\sigma^d$ (where $\Delta {V_b}^c=0$). Therefore, the metastable states go from very long-lived to short-lived as the stress increases, so that the decay of metastability produces the desired negative compressibility transition in an accessible amount time.  Given the constituents needed to achieve negative compressibility transitions, their design may be more easily accomplished using mesoscopic structures, which  will generally involve yet larger energy barriers and hence even more robust metastable states.

Taken together, our results show that one can design metamaterials that exhibit 
both negative compressibility and force amplification transitions (Fig.\ \ref{fig4}). The same properties can be replicated for different network and lattice structures and in three dimensions. 
Potential applications of these materials include
the development of new actuators and protective mechanical devices, and the enhancement of MEMS technologies. Macroscopic implementation 
is achievable with established manufacturing techniques and could  find use in applications such as the design of smart deployable space structures \cite{Mallikarachchi2011}.
We also note that the mechanisms by which mechanical networks can expand in response to pressure and contract in response to tension admit formally equivalent constructions in other types of networks. We anticipate, in particular, that they can be exploited to create new devices with negative finite-difference resistance in electrical and microfluidic networks.

\section*{Methods Summary} \label{ssec.methods}

\noindent
{\it Stable equilibrium of constituents}. For a given $F$, the equilibrium conditions for which all forces are balanced are given by ${\partial V}/{\partial x}= {\partial V}/{\partial y}={\partial V}/{\partial h}= 0$. If $V_x''(x)>0$, $V_y''(y)>0$, and $V_h''(h)>0$, 
as assumed throughout, the equilibrium points $(x^*(F),y^*(F),h^*(F))$ satisfy $h^*=x^*+y^*$. An equilibrium point is linearly stable if the matrix of second derivatives of $V$, denoted $D^2V$, is positive definite at that point. In addition, at stable equilibrium points $\partial h^*/\partial F>0$, confirming that negative compressibility is ruled out for infinitesimal force changes. 

\medskip

\noindent
{\it Condition for destabilisation of constituents}. The Sylvester's criterion applied to the Hessian matrix $D^2V$ shows that an equilibrium point is stable if and only if
\vspace{-0.10cm}
\begin{equation}
V_z^{''}(y^*-x^*)+\frac{2V_x^{''}(x^*)V_y^{''}(y^*)+V_h''(x^*+y^*)\left(V_x''(x^*)+V_y''(y^*)\right)}{V_x^{''}(x^*)+V_y^{''}(y^*)+2V_h''(x^*+y^*)}>0.
\label{mec2}
\vspace{-0.10cm}
\end{equation}
(for details, see Supplementary Information). Destabilisations occur at degenerate equilibrium points, where the inequality is saturated.

\medskip

\noindent
{\it Potentials leading to negative compressibility}.  We let $V_z$ be attractive for all but small $z$ and become negligible for large $z$, 
and we let $V_x$, $V_y$, and $V_h$ be increasingly attractive for large $x$, $y$, and $h$, respectively. 
Unless otherwise noted,
we considered  $V_z$ to be the Lennard-Jones potential, $V_h$ to be a harmonic potential,  and, for conceptual simplicity, $V_x$ and $V_y$ to be piecewise harmonic potentials. The choice of parameters is described in the Supplementary Information, where we also show that similar effects can be obtained using smooth $V_x$ and $V_y$  potentials.

\medskip

\noindent
{\it Molecular dynamics simulations}.  The  simulations were carried out  on a square ``$N$-cell" consisting of $N\times N$ constituents 
 with periodic boundary conditions, and were based on a generalisation of the Andersen barostat \cite{Andersen80} (Supplementary Information). The prescribed strain is achieved by changing the position of the boundaries and the prescribed stress is achieved  by  introducing additional degrees of freedom that account for volume fluctuations.  Constant temperature is achieved using Nos\'e-Hoover thermostat chains \cite{Martyna92}, in which yet additional degrees of freedom are introduced to exchange energy with the system.

\bigskip

\bigskip
\noindent
{\bf Acknowledgements}\,  
This study was supported by the Materials Research Science and Engineering Center at Northwestern University through Grant No.\ DMR-0520513 (Z.G.N.), the National Science Foundation Grants No.\ DMS-0709212 (Z.G.N.\ and A.E.M.) and No.\ DMS-1057128 (A.E.M.), a National Science Foundation Graduate Research Fellowship (Z.G.N.), and a Sloan Research Fellowship (A.E.M.).

\bigskip
\noindent
{\bf Author Contributions}\,  Z.G.N. and A.E.M. designed the research.  Z.G.N. performed the numerical and 
analytical calculations. A.E.M. supervised the research and the analysis of the results. Both authors contributed to the preparation of the manuscript.

\bigskip
\noindent
{\bf Additional information}\, The authors declare no competing financial interests. Supplementary information
accompanies this paper on www.nature.com/nmat/journal/v11/n7/abs/nmat3331.html. Reprints and permissions
information is available online at www.nature.com/reprints. Correspondence and
requests for materials should be addressed to A.E.M.

\newpage

\setcounter{figure}{0}
\renewcommand{\thefigure}{\arabic{figure}}
\renewcommand\figurename{{\bf Fig.}}

\begin{figure*}[!ht] 
\begin{center} 
\raisebox{-5.2cm}{\includegraphics[width=0.8\columnwidth]{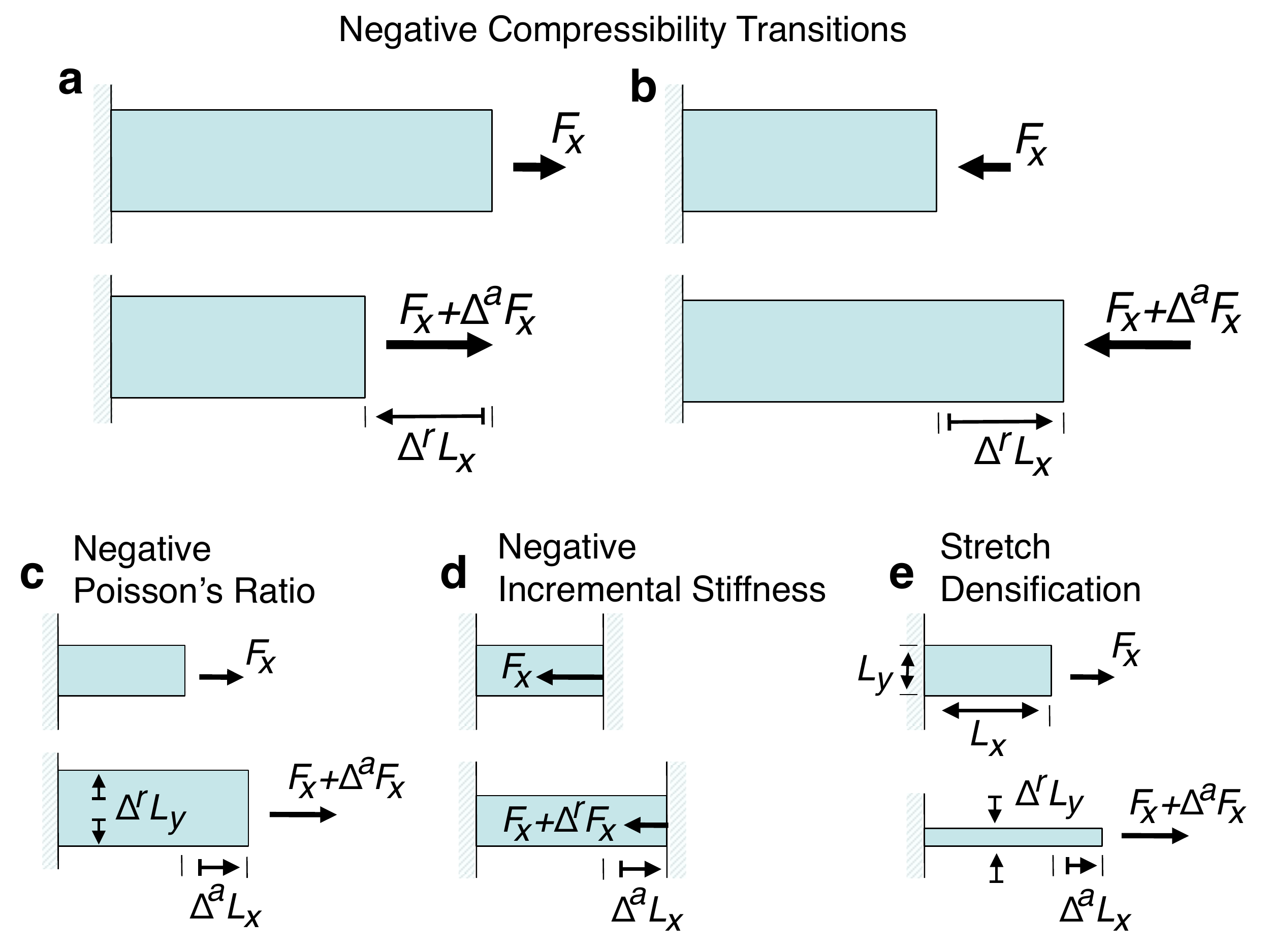}} 
\caption {\baselineskip 14pt
{\bf Negative compressibility contrasted with other effects.} 
{\bf a}, {\bf b}, Negative compressibility transitions: longitudinal deformation $\Delta^{\! \mbox{\scriptsize\it\textsf r}} L_x$ opposing the longitudinal applied force change $\Delta^{\! \mbox{\scriptsize\it\textsf a}} F_x$, which can be tension ({\bf a}) or pressure ({\bf b}), such that $ \Delta^{\! \mbox{\scriptsize\it\textsf r}} L_x/\Delta^{\! \mbox{\scriptsize\it\textsf a}} F_x  <0$.
Here, index {\it\textsf a}\, denotes ``applied," index {\it\textsf r}\, denotes ``response," and  $\Delta^{\! \mbox{\scriptsize\it\textsf a},\mbox{\scriptsize\it\textsf r}}$ is  positive (negative) for an increase (decrease) of the corresponding quantity. 
{\bf c}, Negative Poisson's ratio: transverse expansion (contraction) $\Delta^{\! \mbox{\scriptsize\it\textsf r}} L_y$ in response to longitudinal stretching (shrinking) $\Delta^{\! \mbox{\scriptsize\it\textsf a}} L_x$, such that $- \Delta^{\! \mbox{\scriptsize\it\textsf r}} L_y/\Delta^{\! \mbox{\scriptsize\it\textsf a}} L_x <0$.
{\bf d}, Negative incremental stiffness: decrease in the resulting restoration force $F_x$ in response to an increase in deformation, $\Delta^{\! \mbox{\scriptsize\it\textsf a}} L_x$, such that $\Delta^{\! \mbox{\scriptsize\it\textsf r}} F_x/\Delta^{\! \mbox{\scriptsize\it\textsf a}} L_x <0$. This behaviour can be stable for one component of a composite (e.g., for inclusions in a metal matrix) but not for the material as a whole.
{\bf e},  Stretch-densified materials: materials (with positive bulk modulus) that expand in one direction when hydrostatically compressed also become denser when stretched along this direction, such that  for small deformations  $\Delta^{\! \mbox{\scriptsize\it\textsf a}} L_x$ one has $(\Delta^{\! \mbox{\scriptsize\it\textsf r}}L_y) L_x + (\Delta^{\! \mbox{\scriptsize\it\textsf a}}L_x) L_y <0$. 
Note that the responses in (a) and (b) do not follow from the response in (d). The significant difference is that in (d), due to constraints necessary to stabilise otherwise unstable configurations, the independent variable can be taken to be the deformation $\Delta^{\! \mbox{\scriptsize\it\textsf a}} L_x$  rather than an applied  force  (see also Supplementary Information, Sec.\ 9). This distinction is crucial and should be contrasted with (c) and (e), where it generally makes no difference whether the applied quantity is the longitudinal force or the longitudinal deformation. 
}
\label{fig1}
\end{center}
\end{figure*}

\begin{figure*}[!ht] 
\begin{center} 
\raisebox{-5.2cm}{\includegraphics[width=0.8\columnwidth]{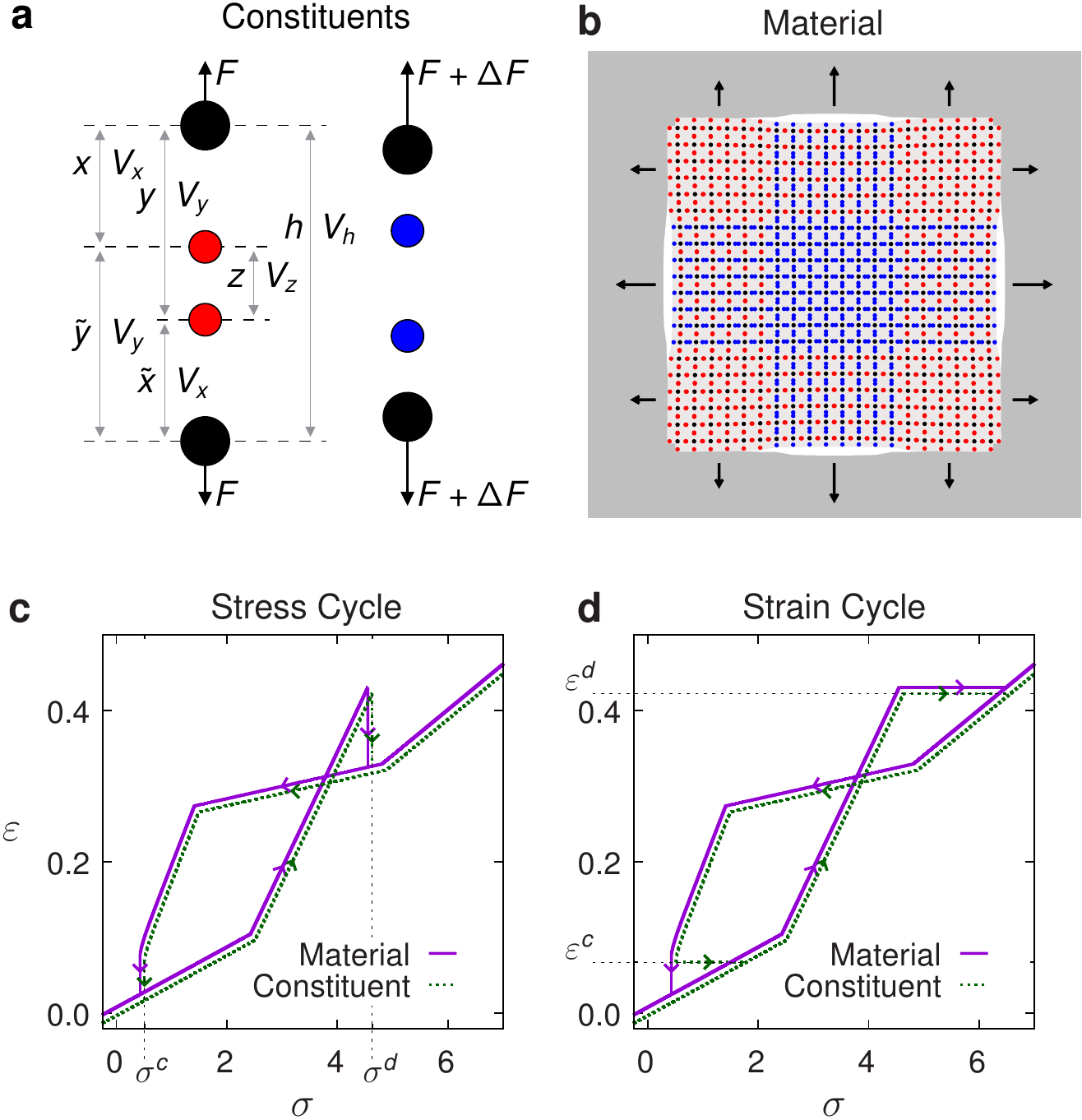}} 
\caption {\baselineskip 14pt
{\bf Constituents, metamaterial, and resulting hysteresis loops.} 
{\bf a}, Constituents, consisting of four particles that interact via potentials $V_x$, $V_y$,  $V_z$, and $V_h$, can be found to be either in a coupled (red) or decoupled (blue) stable configuration. The realised configuration depends on the applied force and history. Note that $z=y-x$, $\tilde{x} = h-y$, and $\tilde{y}=h-x$, so that it takes just three independent variables, $x$, $y$, and $h$, to describe the system completely (see equation (1)).
The system undergoes a negative compressibility transition as the applied force $F$ increases and the configuration changes from coupled (left) to decoupled (right).
{\bf b},  Metamaterial consisting of a square lattice of constituents under a general application of stress, where the individual particles are indicated using the colour scheme of (a). 
There is a small but noticeable contraction (of $\approx 5\%$) at the centre of the sample, where the tension is higher (the force gradient is exaggerated to facilitate visualisation).
The white background marks the boundaries of the material prior to the contraction.
{\bf c}, {\bf d}, Hysteresis loops of the constituents (dotted green) and metamaterial (continuous purple) for tunable uniform applied tensional stress $\sigma$ ({\bf c}), and for tunable uniform applied strain (deformation) $\varepsilon$ ({\bf d}).  For vanishingly small temperature, as considered in this example, the loops of the material and constituents differ only at the coupling destabilisation of strain cycles, which is discontinuous for each constituent and continuous for the material in the thermodynamic  limit. In (c) and (d), the small displacements between the curves were introduced to facilitate visualisation. Because the equilibria of the constituents involve two rather than one degree of freedom, the crossing points of the loops correspond to different equilibria configurations,  one coupled and the other decoupled  (Supplementary Information, Fig.\ S1). 
}
\label{fig2}
\end{center}
\end{figure*}

\begin{figure*}[!ht] 
\begin{center} 
\raisebox{-5.2cm}{\includegraphics[width=0.9\columnwidth]{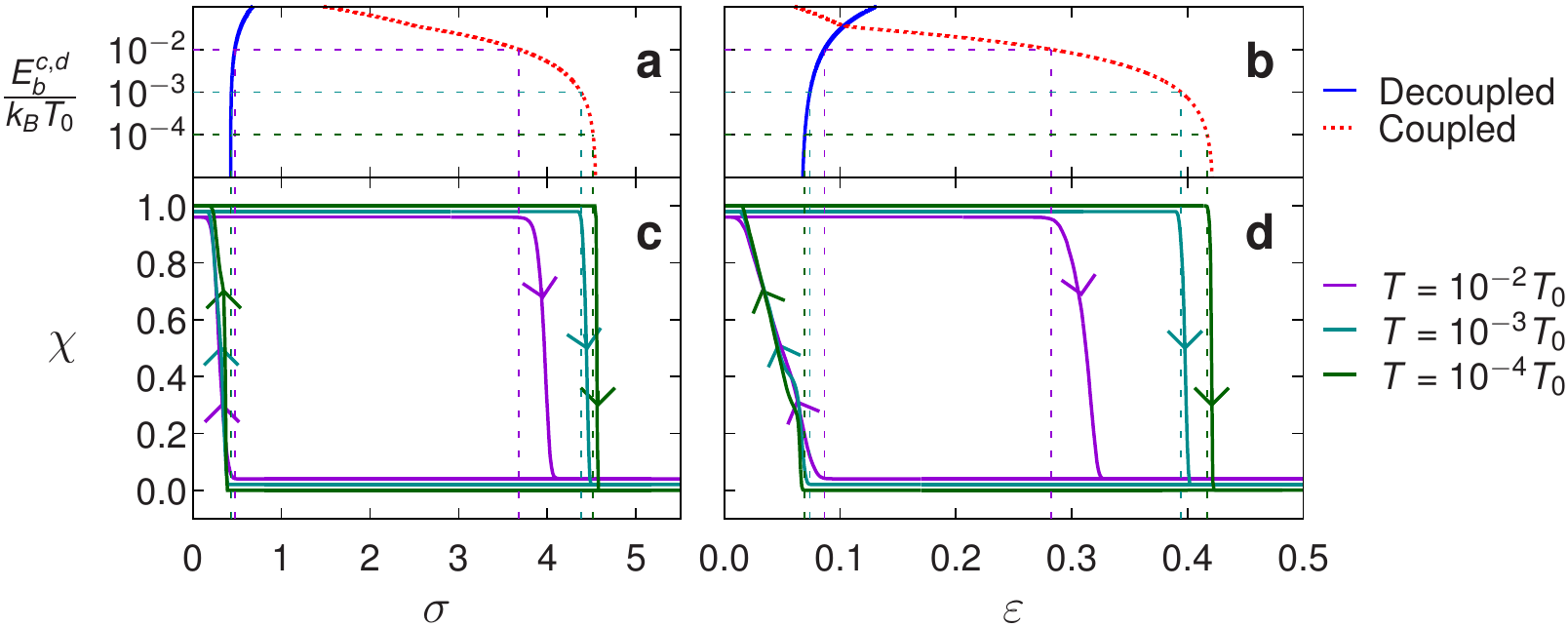}}
\caption {\baselineskip 14pt
{\bf Effective energy barrier and the temperature dependence of the transitions.}  
{\bf a}, {\bf b}, Effective energy barrier between the coupled and decoupled configurations of the constituents, $E^{c,d}_b$,  as a function of the applied stress ({\bf a}) and applied strain ({\bf b}). The height of the barrier is measured relative to the stable coupled state (dotted red; $E^{c}_b$) and to the stable decoupled state (continuous blue; $E^{d}_b$).
{\bf c}, {\bf d}, Fraction of the material in the coupled phase, $\chi$, for several temperatures as a function of the applied stress ({\bf c}) and  applied strain ({\bf d}). 
The curves were generated by averaging over $10$ realisations of our molecular dynamics simulations on a square lattice of $20\times 20$ constituents  
with constant rate of change of the tunable parameter ($\sigma$ or $\varepsilon$) for a cycling time $\tau=10^{1.5}\tau_V$, where $\tau_V$ is the time scale of volume fluctuations. The small vertical displacements in the curves in (c) and (d) were were introduced to facilitate visualisation.  The temperature $T_0$ is a reference, for which
$k_BT_0$ is the potential barrier separating the coupled and decoupled phases when they have the same potential energy. 
The horizontal dotted lines indicate the points where the energy barrier equals the thermal energy $k_BT$. It is observed that these points are excellent predictors of the onset of all coupling and decoupling transitions in (c) and (d) (vertical dotted lines). 
For details on the effective energy barrier,  reference temperature, and volume fluctuation time scale,  see Supplementary Information.
}
\label{fig3}
\end{center}
\end{figure*}

\begin{figure*}[!ht] 
\begin{center} 
\raisebox{-5.2cm}{\includegraphics[width=0.7\columnwidth]{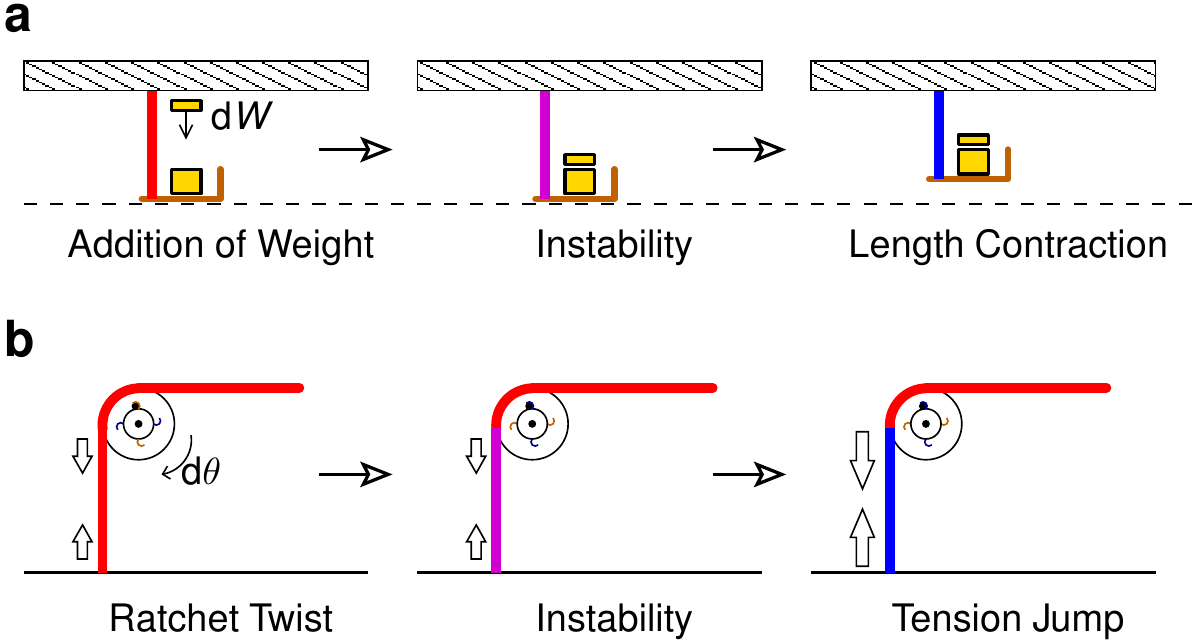}}
\caption {\baselineskip 14pt
{\bf Physical examples of stress-induced negative compressibility and strain-induced force amplification transitions.} 
{\bf a}, A rope made of our material holds a basket containing a weight, which stretches the rope to near the decoupling transition point.  When a small amount of extra weight $\mathrm{d}W$ is added to the basket, an instability develops, and the rope spontaneously passes from the coupled to the decoupled phase.  As a result, it raises the basket, despite the increase in the weight.
{\bf b}, A rope made of our material is attached to the ground and to a raised ratchet. The rope is stretched to a length near the decoupling transition point.  When the ratchet is rotated a small amount $\mathrm{d}\theta$, an instability causes the rope to pass from the coupled to the decoupled phase. As a result, the tension in the rope is amplified by a large amount. 
}
\label{fig4}
\end{center}
\end{figure*}


\begin{thebibliography}{99}

\baselineskip 15pt

\bibitem{Veselago68}
Veselago, V. G.
The electrodynamics of substances with simultaneously negative values of $\varepsilon$ and $\mu$. 
{\it Sov. Phys. Usp.} {\bf 10}, 509--514 (1968). Original Publication in Russian, 1967.

\bibitem{Smith00}
Smith, D. R., Padilla, W. J., Vier, D. C., Nemat-Nasser, S. C. \&  Schultz,  S.
Composite medium with simultaneously negative permeability and permittivity.
{\it Phys. Rev. Lett.} {\bf 84}, 4184--4187 (2000).

\bibitem{Shelby01}
Shelby, R. A., Smith, D. R. \& Schultz, S.
Experimental verification of a negative index of refraction.
{\it Science} {\bf  292}, 77--79 (2001).

\bibitem{Pendry00}
Pendry, J. B.
Negative refraction makes a perfect lens.
{\it Phys. Rev. Lett.}  {\bf 85}, 3966--3969 (2000).

\bibitem{Schurig06}
Schurig, D., {\it et al.}
Metamaterial electromagnetic cloak at microwave frequencies. 
{\it Science} {\bf 314}, 977--980 (2006).

\bibitem{Zhang09}
Zhang, S., Yin, L. \&  Fang, N.  
Focusing ultrasound with an acoustic metamaterial network.
{\it Phys. Rev. Lett.} {\bf 102}, 194301 (2009).

\bibitem{Baughman98}
Baughman, R. H., Stafstr\"om, S., Cui, C. \& Dantas, S. O.  
Materials with negative compressibilities in one or more dimensions. 
{\it Science} {\bf 279}, 1522--1524 (1998).

\bibitem{Lee02}
Lee, Y., Vogt, T., Hriljac, J. A., Parise, J. B. \& Artioli, G.
Pressure-induced volume expansion of zeolites in the natrolite family.
{\it J. Am. Chem. Soc.} {\bf 124}, 5466--5475 (2002).

\bibitem{Vakarin06}
Vakarin, E. V., Duda, Y. \& Badiali, J. P. 
Negative linear compressibility in confined dilatating systems.
{\it J. Chem. Phys.} {\bf 124}, 144515 (2006).

\bibitem{grima08}	
Grima, J. N.,  Attard, D. \&  Gatt, R.
Truss-type systems exhibiting negative compressibility.
{\it Phys. Stat. Sol. (B)} {\bf 245}, 2405--2414 (2008).

\bibitem{gatt08} 
Gatt, R. \& Grima, J. N.
Negative compressibility.
{\it Phys. Stat. Sol. (RRL)} {\bf 2}, 236--238 (2008). 

\bibitem{Liu00}
Liu, Z., {\it et al.} % 7 authors
Locally resonant sonic materials. 
{\it Science} {\bf 289}, 1734--1736 (2000).

\bibitem{Fang06}
Fang, N., {\it et al.} % 7 authors
Ultrasonic metamaterials with negative modulus. 
{\it Nat. Mater.} {\bf 5}, 452--456 (2006).

\bibitem{Sheng03}
Sheng, P., Xiao, R.-F., Wen, W.-J. \& Zheng, Y.  L.
Composite materials with negative elastic constants. US Patent 6576333.
Washington, DC: US Patent and Trademark Office (2003).

\bibitem{Lakes01}
Lakes, R. S., Lee, T., Bersie, A. \& Wang, Y. C.
Extreme damping in composite materials with negative-stiffness inclusions.
{\it Nature} {\bf 410}, 565--567 (2001).

\bibitem{jaglinsk07}
Jaglinski, T., Kochmann, D., Stone D. \& Lakes, R. S.
Composite materials with viscoelastic stiffness greater than diamond.
{\it Science}  {\bf 315}, 620--622 (2007).

\bibitem{Reichl2009}
Reichl, L. E. {\em A Modern Course in Statistical Physics} (Wiley-VCH, New York, NY, 2009).

\bibitem{Lakes87}
Lakes, R. S. 
Foam structures with a negative Poisson's ratio. 
{\it Science} {\bf 235}, 1038--1040 (1987).

\bibitem{Baughman98-2}
Baughman, R. H., Shacklette, J. M., Zakhidov, A. A. \& Stafstr\"{o}m, S. 
Negative Poisson's ratio as a common feature of cubic metals. 
{\it Nature} {\bf 392}, 362--365 (1998).

\bibitem{Janmey07}
Janmey, P. A., McCormick, M. E., Rammensee, S., Leight, J. L., Georges, P. C. \& MacKintosh, F. C.
Negative normal stress in semiflexible biopolymer gels.
{\it Nat. Mater.} {\bf 6}, 48--51 (2007).

\bibitem{Moore06}
Moore, B., Jaglinski, T., Stone, D. S. \& Lakes, R. S.
Negative incremental bulk modulus in foams.
{\it Phil. Mag. Lett.} {\bf 86}, 651--659 (2006).

\bibitem{Goodwin08}
Goodwin, A. L., Keen, D. A. \& Tucker, M. G.
Large negative linear compressibility of Ag$_3$[Co(CN)$_6$].
{\it Proc. Natl. Acad. Sci. USA} {\bf 105}, 18708--18713 (2008).

\bibitem{Fortes2011}
Fortes, D. A., Suard, E. \&  Knight, K. S.
Negative linear compressibility and massive anisotropic thermal expansion in methanol monohydrate.
{\it Science}  {\bf 331}, 742--746 (2011). 

\bibitem{Braess05} 
Braess, D., Nagurney, A. \& Wakolbinger, T. 
On a paradox of traffic planning. {\it Transp. Sci.} {\bf 39}, 
446--450 (2005). Original Publication in German, 1968.

\bibitem{Roughgarden05}
Roughgarden, T.
{\it Selfish Routing and the Price of Anarchy} 
(MIT Press, Cambridge, MA, 2005). 

 \bibitem{Beckmann56}
Beckmann, M. J., McGuire, C. B. \& Winsten, C. B.
{\it Studies in the Economics of Transportation} 
(Yale Univ. Press, New Haven, CT, 1956).  

\bibitem{Pigou20}
Pigou, A. C. 
{\it The Economics of Welfare} 
(Macmillan and Co., London, UK, 1920). 

\bibitem{Cohen91}
Cohen, J. E. \& Horowitz, P. 
Paradoxical behaviour of mechanical and electrical networks. 
{\it Nature} {\bf 352}, 699--701 (1991).

\bibitem{stanley}
Mishima, O. \& Stanley,  H. E.
The relationship between liquid, supercooled and glassy water.
{\it Nature} {\bf 396}, 329--335 (1998).

\bibitem{otsuka}
Otsuka, K. \& Wayman, C. M. \em Shape Memory Materials \em (Cambridge Univ. Press, New York, NY, 1998).

\bibitem{abeyaratne}
Abeyaratne, R. \& Knowles, J. K. \em Evolution of Phase Transitions: A Continuum Theory \em (Cambridge Univ. Press, New York, NY, 2006).

\bibitem{Puglisi2000}
Puglisi, G. \&  Truskinovsky, L. 
Mechanics of a discrete chain with bi-stable elements. 
{\it J.  Mech. Phys. Solids}  {\bf 48}, 1--27 (2000).

\bibitem{Mallikarachchi2011}
Mallikarachchi, H. M. Y. C. \& Pellegrino, S. 
Quasi-static folding and deployment of ultrathin composite tape-spring hinges. 
{\it J. Spacecraft Rockets} {\bf 48}, 187--198 (2011).

\bibitem{Andersen80}
Andersen, H. C.
Molecular dynamics simulations at constant pressure and/or temperature.
{\it J. Chem. Phys.} {\bf 72}, 2384--2393 (1980).

\bibitem{Martyna92}
Martyna, G. J., Klein, M. L. \& Tuckerman, M. 
Nos\'{e}-Hoover chains: The canonical ensemble via continuous dynamics. 
{\it J. Chem. Phys.} {\bf 97}, 2635--2643 (1992).


\end{thebibliography}
\end{document}